# First Demonstration of Flip DRAM from Process, Architecture to System to Push DRAM Scaling beyond $4F^2$: $2F^2$ Self-aligned Flip Vertical Channel Transistor (FVCT) DRAM and Flip WL (FWL) 3D-DRAM

Yu Liu[1], Xinyue He[1], Yanbang Chu[1], Siyuan Liu[1], Jianxiang Jin[1], Fangcheng Sun[1,2], Yuyang Qiao[3], Xu Tian[3], Lan Li[4], Baokang Peng[3], Lining Zhang[3], Xing Wu[4], Pengpeng Ren[2], Zhigang Ji[2], Zongwei Wang[1], Lijie Zhang[1], Xinwei Wang[3], Weihai Bu[1], Ming Li[1], Runsheng Wang[1], Heng Wu[1,*], Ru Huang[1]

[1]School of Integrated Circuits, Peking University, Beijing, China (*Email: hengwu@pku.edu.cn ); [2]School of Integrated Circuits, SJTU, Shanghai, China; [3]Peking University Shenzhen Graduate School, Shenzhen, China; [4]School of Integrated Circuits, ECNU, Shanghai, China

***Abstract*—For the first time, we proposed a novel stacking technology for DRAM scaling by flipping and backside processes, making full use of DRAM wafer's backside and investigating it on both $4F^2$ and 3D-DRAM. For $4F^2$ VCT, $2F^2$ Flip VCT featuring self-aligned back-to-back stacked 1T1C bitcell, with various BL & WL configurations, were studied and key process modules such as self-aligned stacked vertical channel, BL & WL formations, wafer bonding & flipping, substrate thinning and low-R Co storage node (SN) were successfully developed, addressing the potential thermal, misalign and parasitic concerns in the Flip VCT process. A full DRAM DTCO framework was also established from device to mat and chip level. Compared to $4F^2$ VCT DRAM with the same mat size, $2F^2$ FVCT delivers 27.5% less parasitics, 11% better sense margin, 16.3% higher charge sharing (CS) speed and 50% less area. For 3D-DRAM, a brand-new flip WL staircase design with peripheral circuit innovations was studied and proved to have 25% density gain, 15.1% faster turn-on speed and 6.8% less CS time, proving further extendibility of flip technology on DRAM.**

## I. Introduction

In recent years, demands for higher density and larger capacity DRAM have increased rapidly with the unprecedented advancement of AI [1]. As the DRAM scales down to sub-10 nm, the lateral $6F^2$ DRAM faces great challenges in process complexity and scalability [2]. By decoupling BL out of the plane of SN, $4F^2$ VCT DRAM achieves smaller cell size and less parasitics [3], confirmed as the next solution on the roadmap as in Fig. 1 shows. However, the extendibility of $4F^2$ VCT is still questionable for the high cost of non-sharable processes with $6F^2$ and future's 3D-DRAM [4].

To extend the lifespan of $4F^2$ VCT DRAM, we first proposed and investigated a novel $2F^2$ FVCT architecture (Fig. 1), featuring self-aligned back-to-back stacked VCT cells to fully utilize the wafer backside (BS). Key processes of vertical nanosheet, bonding, flipping, substrate strip and self-aligned BL/WL were successfully developed , fully compatible with standard $4F^2$ processes. Other key $2F^2$ DRAM alternatives with different BL/WL schemes were also benchmarked from circuit to chip level. Furthermore, by placing orthogonal VCTs and hexagon caps [5], ultra scaled $1.75F^2$ FVCT is proposed.

In addition, 3D-DRAM is also widely discussed for its multi-stacking nature, especially the vertical BL (VBL) approach with low BL parasitics [6]. However, the resulting horizontal WL (HWL) requires very area-consuming staircases connection, diminishing the density gain [7,8]. Unlike 3D-NAND with kBs cells per WL [9], DRAM urges for speed thus each WL connects much fewer cells, resulting in even larger staircase footprint per cell [10]. Fortunately, it can be effectively mitigated by flipping WL staircases by half to the BS, as first-time proposed and evaluated in this work (Fig. 1). By adjusting the peripheral circuits to match the WL connections, FWL 3D-DRAM delivers not only clear density gain but also higher performance from the much reduced WL parasitics.

This work was supported in part by the National Key R&D Program of China: Grant 2023YFB4402201; in part by NSFC: Grant 92464206.

## II. $2F^2$ FVCT DRAM Processes and Optimizations

FVCT can double the cell density and the Frontside (FS) & BS VCTs are connected at the VCT bottom by the back-to-back stacking, featuring three distinct BL & WL configurations as in Fig. 1: 1) the baseline **c**BL**s**WL (common BL, split WL) with FS and BS cells sharing the same BL but WLs separated; two alternatives: 2) **s**BL**s**WL with split WL and split BL for FS & BS VCTs; 3) **s**BL**c**WL with common WL but split BL for FS & BS VCTs.

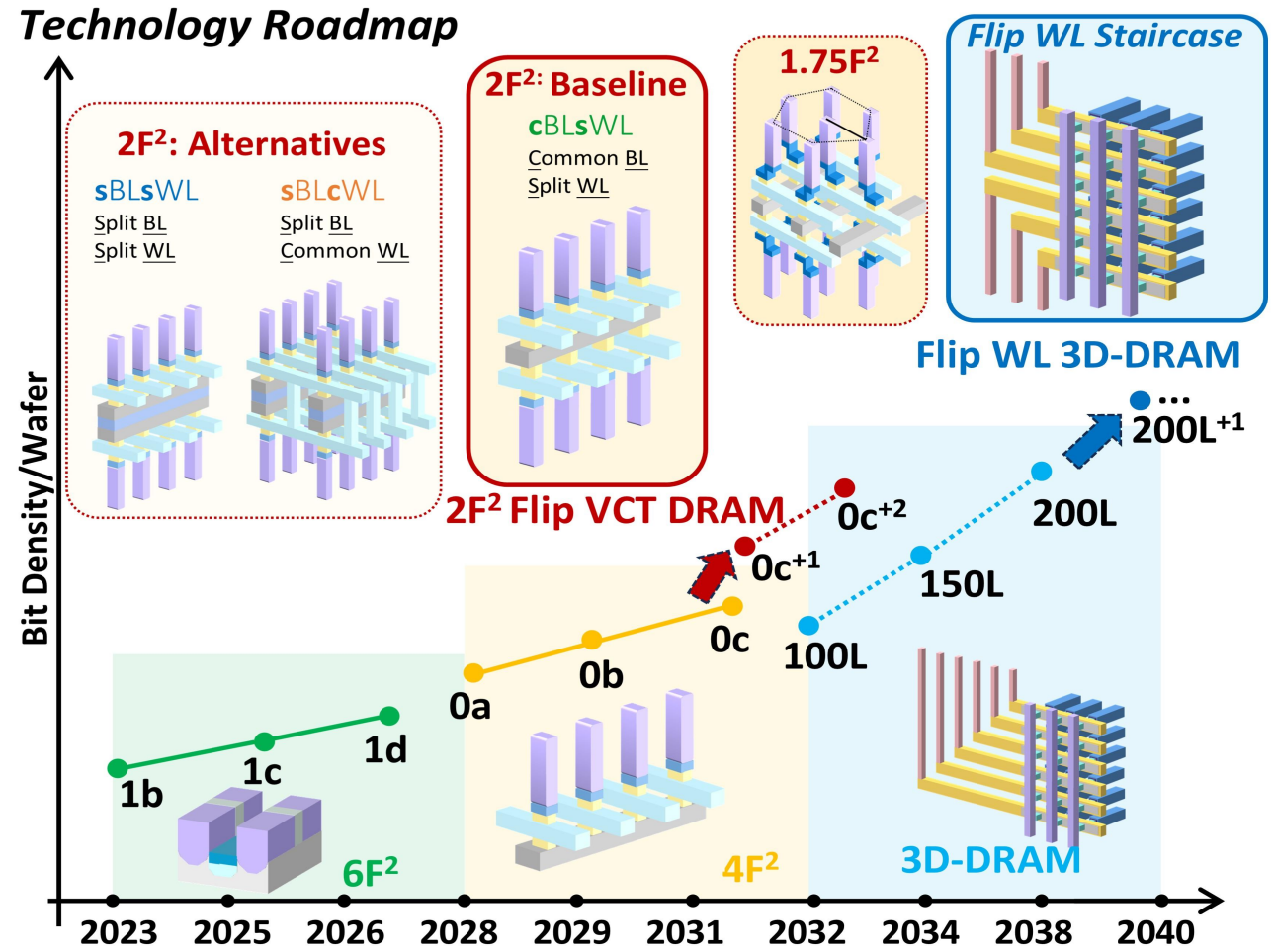

Fig. 1. The roadmap of DRAM scaling. To increase the bit density, for $4F^2$, $2F^2$ Flip VCT DRAM was proposed with different architectures; for 3D-DRAM, flip WL 3D-DRAM was introduced.

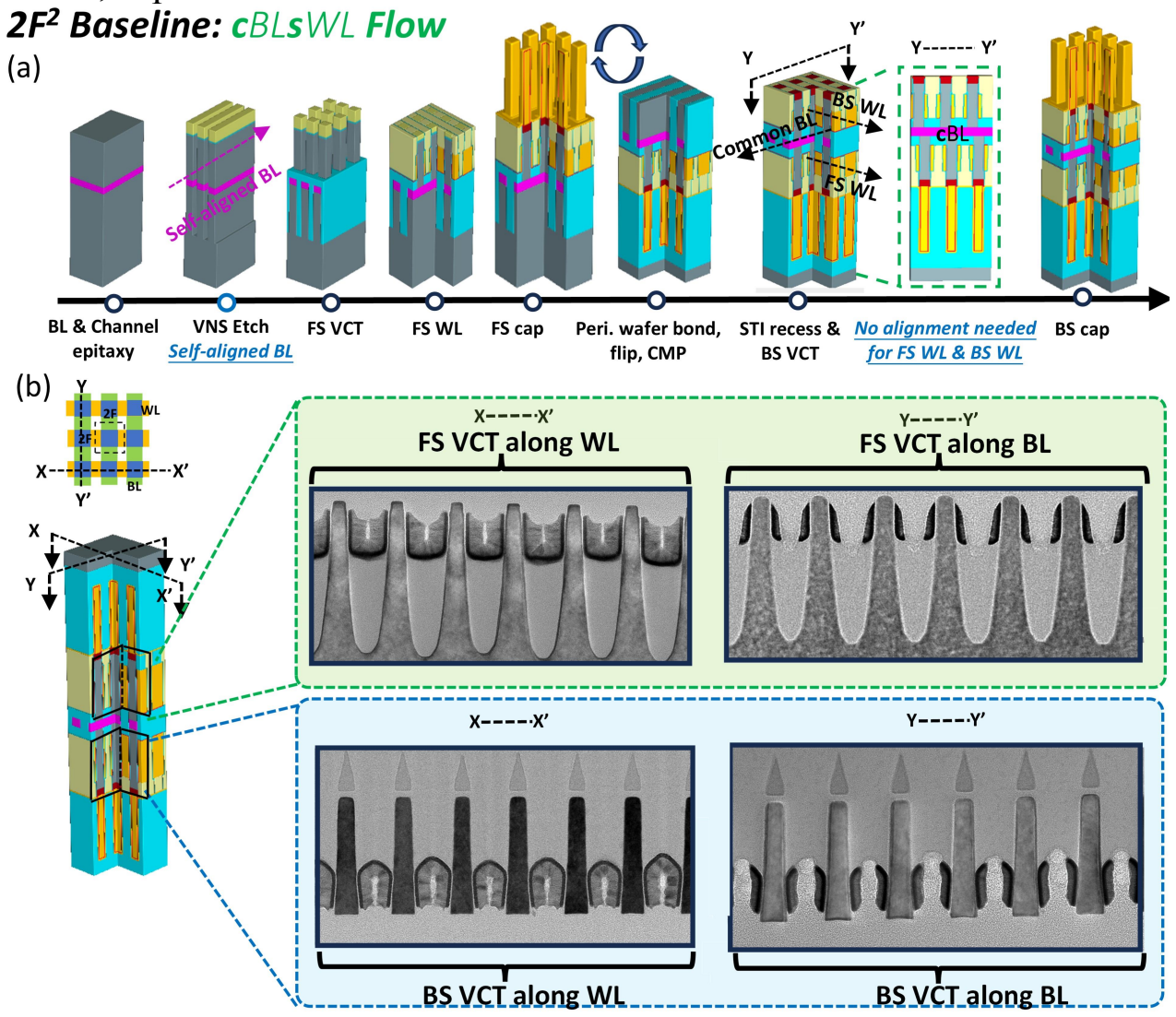

Fig. 2. $2F^2$ baseline flow (**c**BL**s**WL). (a) FVCT with self-aligned BL and no alignment needed for FS & BS WLs. (b) TEMs for FS & BS VCT WLs. Wafer bonding, flipping & substrate removal were developed and utilized in the formation of BS VCT WLs.

## *A. Baseline Processes*

Fig. 2(a) shows the proposed baseline process flow of **c**BL**s**WL. After BL & vertical channel epitaxy, the vertical nanosheet (VNS) is etched and half buried in STI, forming self-aligned BL, followed by the standard $4F^2$ process to finish the FS VCT bitcell. Then a carrier or peripheral wafer is bonded and flipped, succeeded by substrate removal and STI recess to expose the BS. Finally, the BS 1T1C is formed. Note that, in FVCT, the FS & BS WLs are formed on both sides of the wafer separately, without requiring aligning the two. Key processes of FS & BS WLs were also developed as in Fig. 2(b), relying on robust wafer bonding, flipping and substrate removal processes. It should be admitted that the FS & BS modules were developed separately for the concept and further work is needed to combine the two.

## *B. Process Optimizations*

Considering the FS capacitor may degrade from the BS VCT's thermal processes in the baseline, a thermal budget friendly flow with cap formation last and self-aligned WL was proposed (Fig. 3). After forming two stacked VCTs & FS caps on the FS monolithically, the wafer is flipped to form the BS caps to protect the FS/BS caps from VCT thermals. Moreover, the novel rectangle layout enables WL pinch-off during ALD gate formation without litho. definition (Fig. 3(b)).

Besides, key modules such as high aspect ratio (1:20) VNS etch & low-resistance Co storage node metallization processes were developed (Figs. 4(a-b)). Two common BL schemes were also proposed (Fig. 4(c)): which are 1) highly doped Silicon BL wrapped around by TiN with relaxed process complexity and 2) metal-only BL for smaller resistance: after self-aligned FS & BS VCTs formation, FS VCT is then suspended at the bottom, followed by TiN BL gap fill and recess.

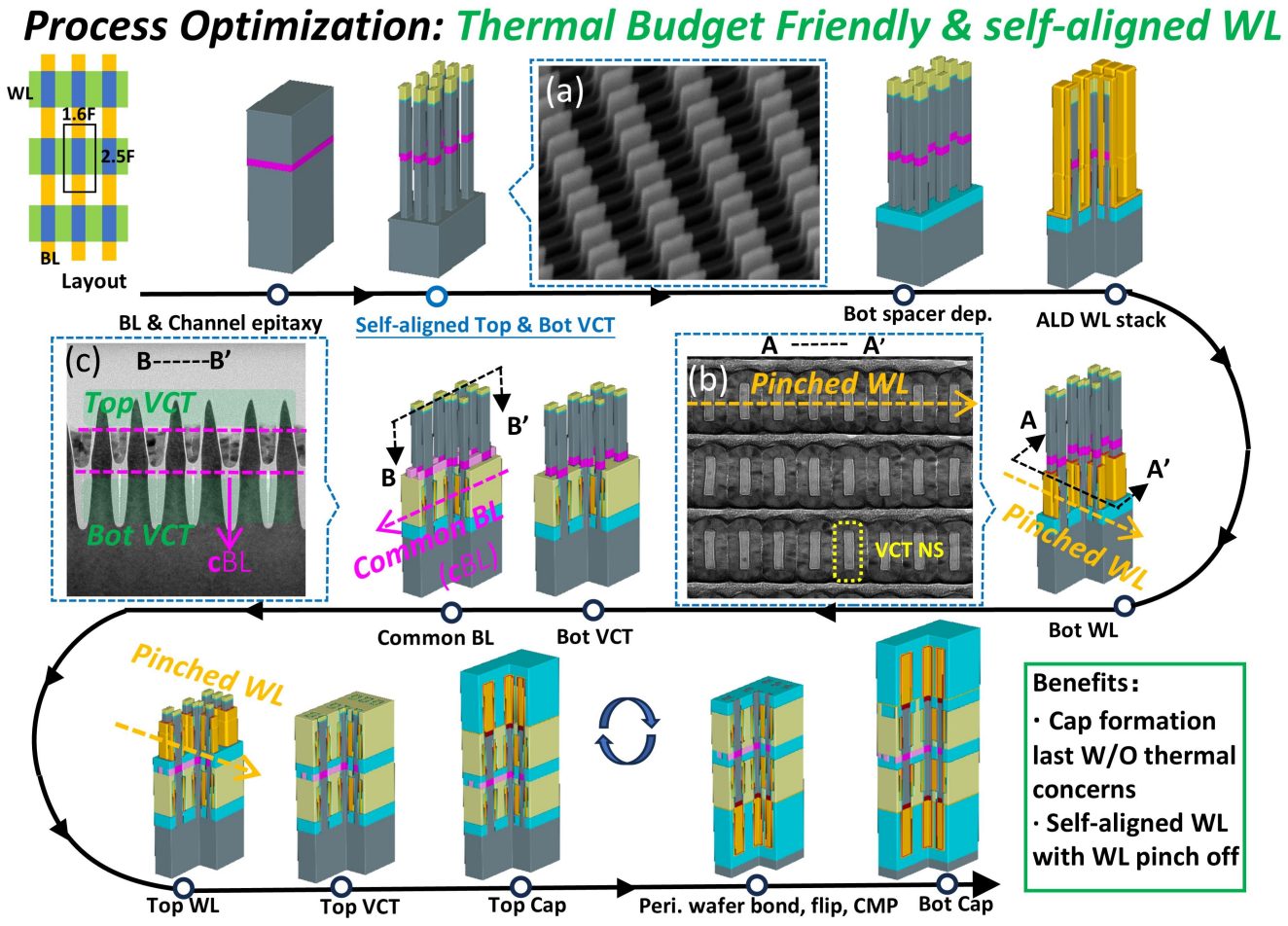

Fig. 3. Process flow optimizations on $2F^2$ for the potential thermal & misalign concerns with capacitor formation last and pinched WL. (a) Stacked VCTs array. (b) Self-aligned WL with WL pinch-off. (c) FS & BS VCTs share the Common BL (**c**BL) .

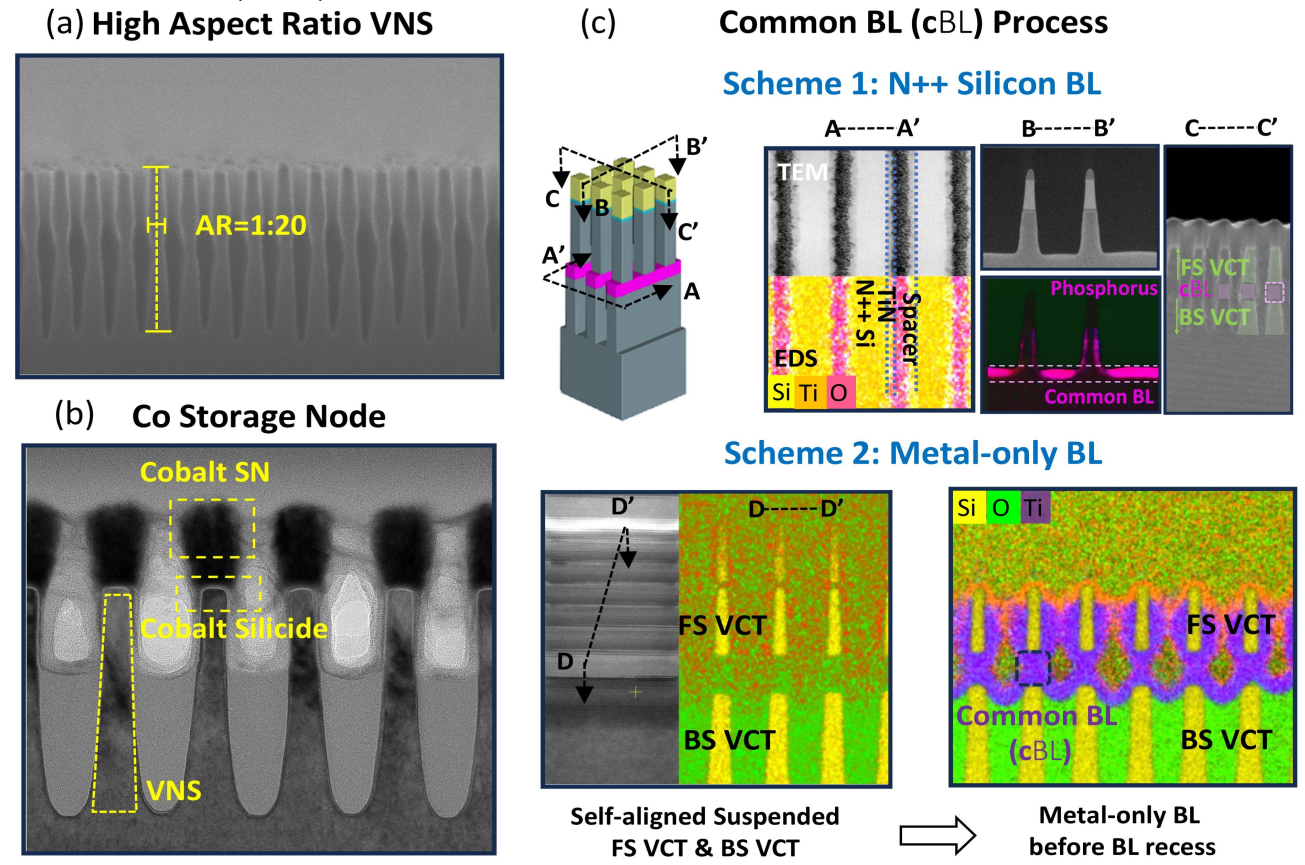

Fig. 4. (a-b) Module process optimizations on HAR etch (a) and Co storage node (b). (c) Two schemes to form **c**BL: highly doped N++ BL covered by TiN and metal BL with TiN filling in the gap between FS & BS VCTs.

## *C. Architecture optimizations*

**s**BL**s**WL & **s**BL**c**WL as two new BL/WL connection architectures were also studied to extend FVCT's flexibility as in Fig. 5. After forming the two separated but self-aligned BLs in the TEM of Fig. 5(a), **s**BL**s**WL's WLs are separated while **s**BL**c**WL adds a side via to connect FS & BS WLs together.

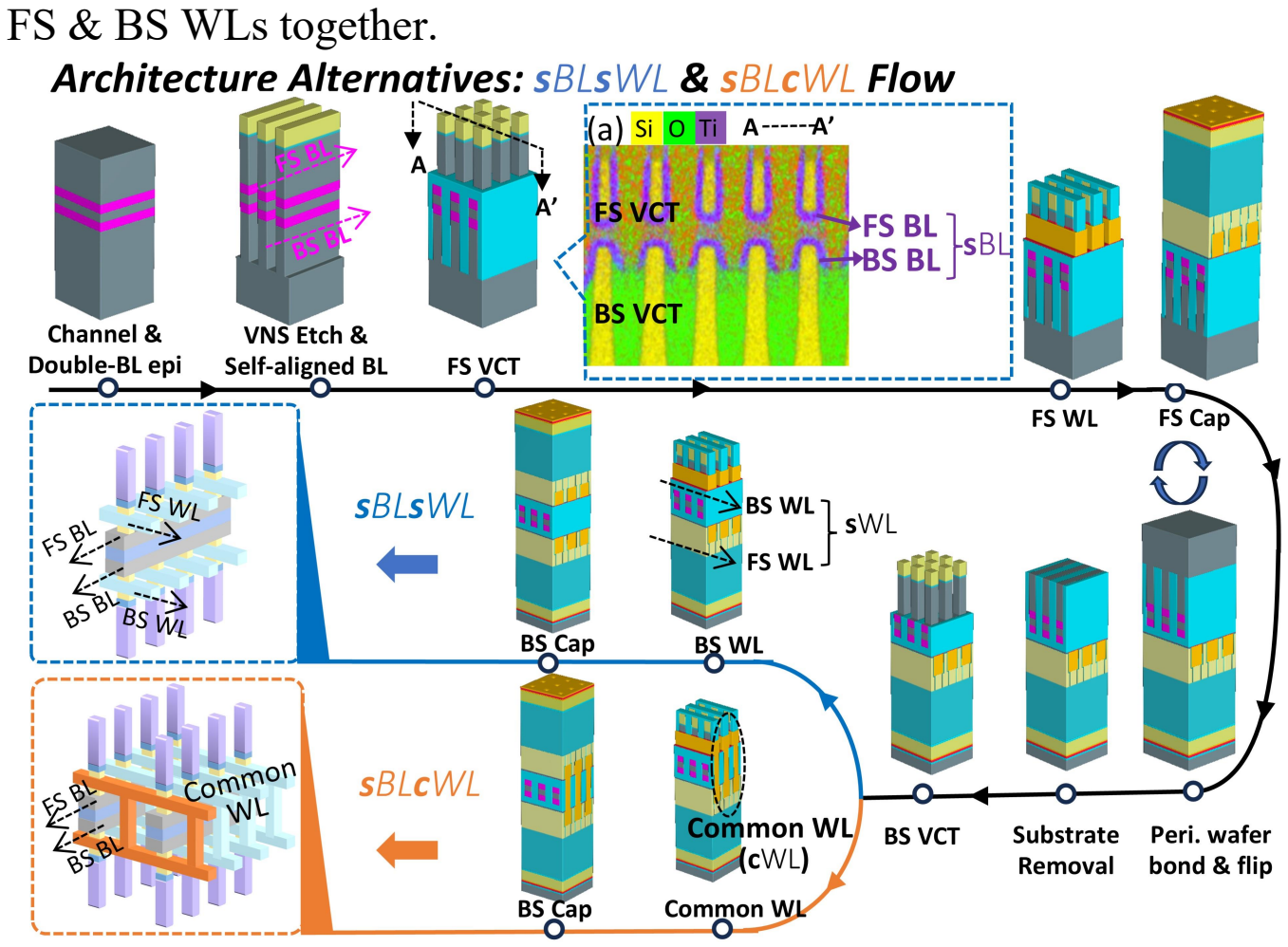

Fig. 5. Flow of the other two FVCT options: **s**BL**s**WL forms separated WLs while **s**BL**c**WL adds a side via to connect FS & BS WLs. (a) TEM of split BL with separated but self-aligned FS & BS BLs.

## III. Circuit and Chip Benchmark on $2F^2$ FVCT

Calibrated to industry data [2], the VCT's SPICE model was fitted by the neural network & deep learning [11] with error<0.95% (Fig. 6(a)). RC netlists of various DRAM array sizes were first extracted by process emulation & parasitic extraction, then fitted and extrapolated to a larger array of 1024×1024 to evaluate FVCT architectures, with good matching results as in Figs. 6(b-c).

## *A. Parasitics comparison*

As in Fig. 6(d), $2F^2$ FVCT has a larger $C_{BL}$ considering both BL parasitics of FS and BS bitcells, among which **c**BL**s**WL has the least increase of 45% (but equivalently 27.5% smaller per cell) than $4F^2$. **s**BL**c**WL has 105% larger $C_{WL}$ (equivalently 2.5% larger per cell) for WL connecting both FS & BS bitcells.

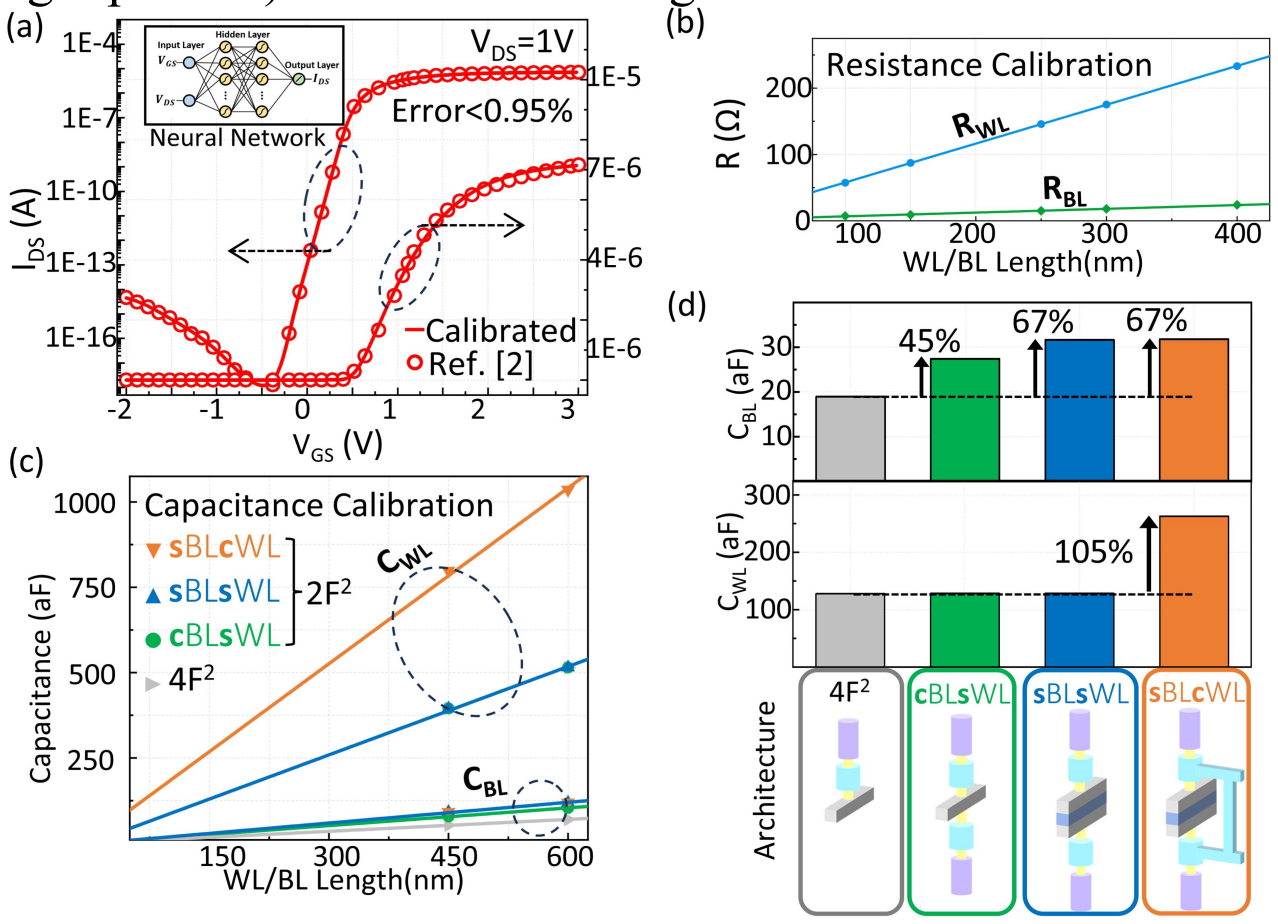

Fig. 6. (a) SPICE model extraction by neural network (error<0.95%). (b-c) RC calibration by fitting RC to various array sizes, showing data match well. (d) BL & WL capacitances for $4F^2$ and $2F^2$ DRAMs.

## *B. Array design*

$2F^2$ FVCT also features clear advantages in mat floorplan: **c**BL**s**WL can fold BLs to achieve 50% BL length reduction and **s**BL**c**WL has shorter WL length for folding WLs. However, **s**BL**s**WL is different (Fig. 7(a)), by interleaving & folding split cells based on the standard folded-BL layout, shrinking pitch of WLs & BLs [12], the array footprint can be reduced to 25% of the original, namely the pseudo folded-BL design.

Furthermore, for array arrangement in a bank, since BLs must be placed at both ends of the sense amplifier (SA), when it comes to the bank edge, only half of the bank edge's BLs can be connected to SA and the other half become dummy BLs [13] in $4F^2$ & **c**BL**s**WL & **s**BL**c**WL (Fig. 7(b)). Unfortunately, these dummy BLs are unavoidable for the process uniformity. However, thanks to the unique split BL & WL design, **s**BL**s**WL has FS & BS BL connected to the same SA, saving dummy BLs area at the bank edge.

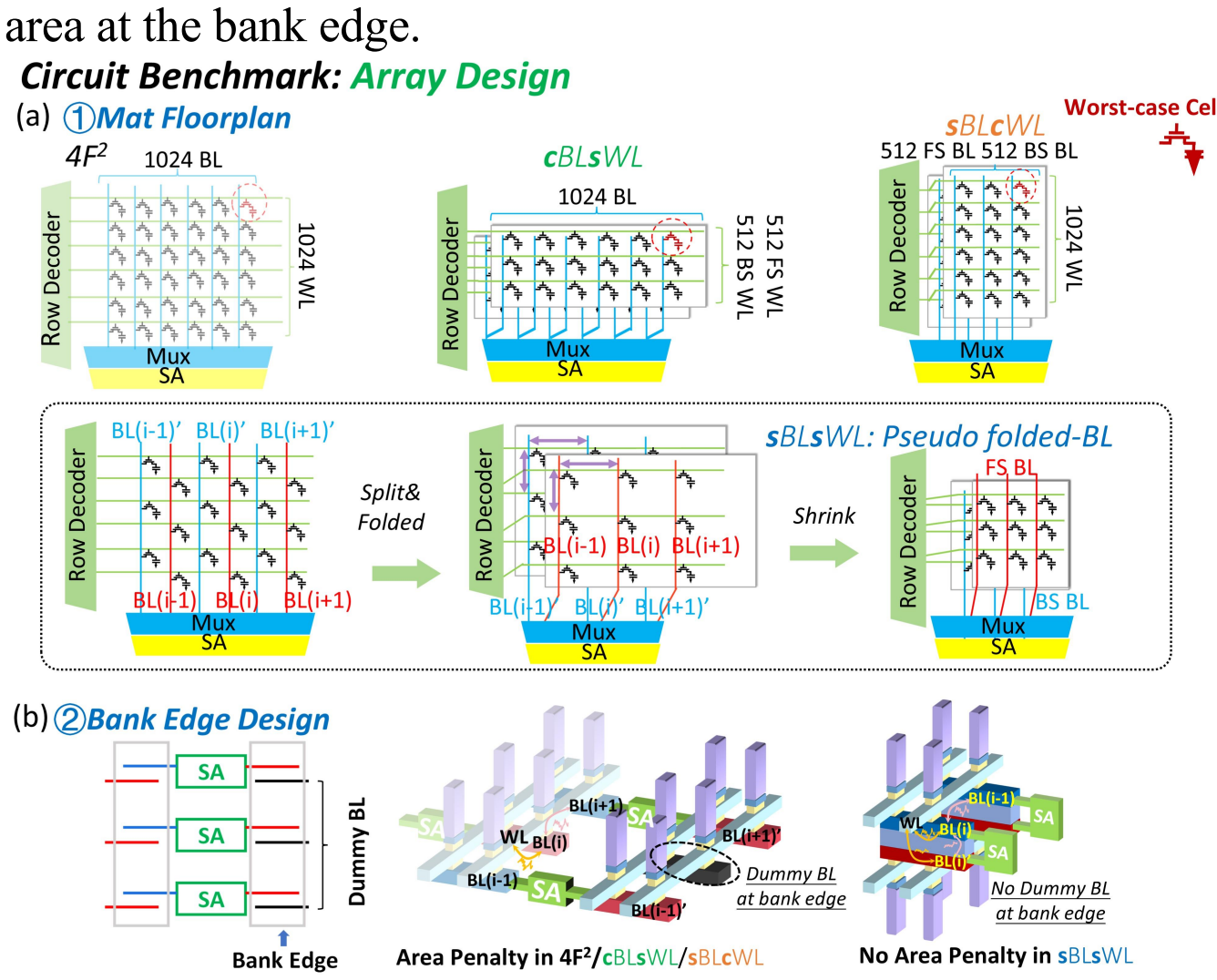

Fig. 7. (a) Mat floorplans of $4F^2$ and $2F^2$ DRAMs with different architecutures. **c**BL**s**WL folds BL by half while **s**BL**c**WL folds WL by half to fully leverage $2F^2$ advantages and achieve smaller area. **s**BL**s**WL features pseudo folded-BL with splitting and folding cells based on the standard folded-BL design. (b) **s**BL**s**WL eliminates the dummy BL at the bank edge, with clear area benefits.

## *C. Noise & Sense margin (SM)*

The peripheral circuits for BL are shown in Fig. 8(a) and used for DRAM operations. The waveforms of the whole read & write are given in Fig. 8(b).

Noise, mainly attributed to the WL & BL coupling on BLs as marked by yellow arrows in Fig.7(b) , is critical to the DRAM's operation, in which the BL-BL coupling noise can be divided into stage 1 & 2, referring to the noises before and after triggering the SA [14], with the waveforms shown in Fig. 8(c).

Due to the different arrival time of SA signals, the BLs in unopened SA can be seriously affected by the opened ones with the amplified signals, which is the stage 2 noise. For pseudo folded-BL & open-BL designs, the worst case is that SN(i) is 1 but its nearby SNs are 0. $4F^2$ & **s**BL**c**WL have prominent stage 2 bit errors (Fig.8 (d)), indicating BL signal is reversed due to the interference from the adjacent BLs from the opened SA region. But **c**BL**s**WL is better due to the smaller parasitics while **s**BL**s**WL is the best for the pseudo folded-BL design, in which the operational BLs experience the same environmental noise as nearby BLs, thus its BL-BL noise can be further optimized by offsetting the common mode components. Meanwhile, the operational BLs are also affected by the coupling of the nearby diagonal BLs so that the noise can be partially compensated, benefiting the stability of the operational BLs.

In addition, as a factor considering the coupling of WL & BL, the R factor was used to evaluate WL-BL noise [15], **c**BL**s**WL exceeds $4F^2$ by 35% (Fig. 8(d)). Fig. 8(e) shows the ideal sense margin, **c**BL**s**WL has the largest sense margin due to the smaller BL parasitics, exceeding $4F^2$ by 11%.

## *D. Read & Write Latency at the chip level*

FVCT was further assessed at the chip level. Key timing metrics on WL delay and charge sharing were extracted (Figs. 9(a-b)). **s**BL**c**WL has 11.5% less WL turn-on delay than $4F^2$ while **c**BL**s**WL reaches stable charge sharing first with the largest SM. These parameters were then fed to Ramulator (Fig. 9(c)) [16], a widely used open-source DRAM simulator, in which DRAM controller is driven by various workload traces and the output latency is given in Fig. 9(d). **c**BL**s**WL performs better under the four traces, **s**BL**c**WL is the best only in memory trace, where row switching operations happen frequently as in database random queries. Table I compares the $4F^2$ & 3 FVCT architectures and the baseline **c**BL**s**WL is overall the best.

Circuit Benchmark: Noise & Sense Margin

Fig. 8. (a) The BL peripheral circuits in the simulation. (b) The read & write waveforms. (c) BL-BL noise divided to stage 1 & 2, referring to the noises before & after triggering the SA. (d) Results of WL-BL & BL-BL noises. (e) Ideal sense margin of $4F^2$ and $2F^2$ FVCT DRAMs.

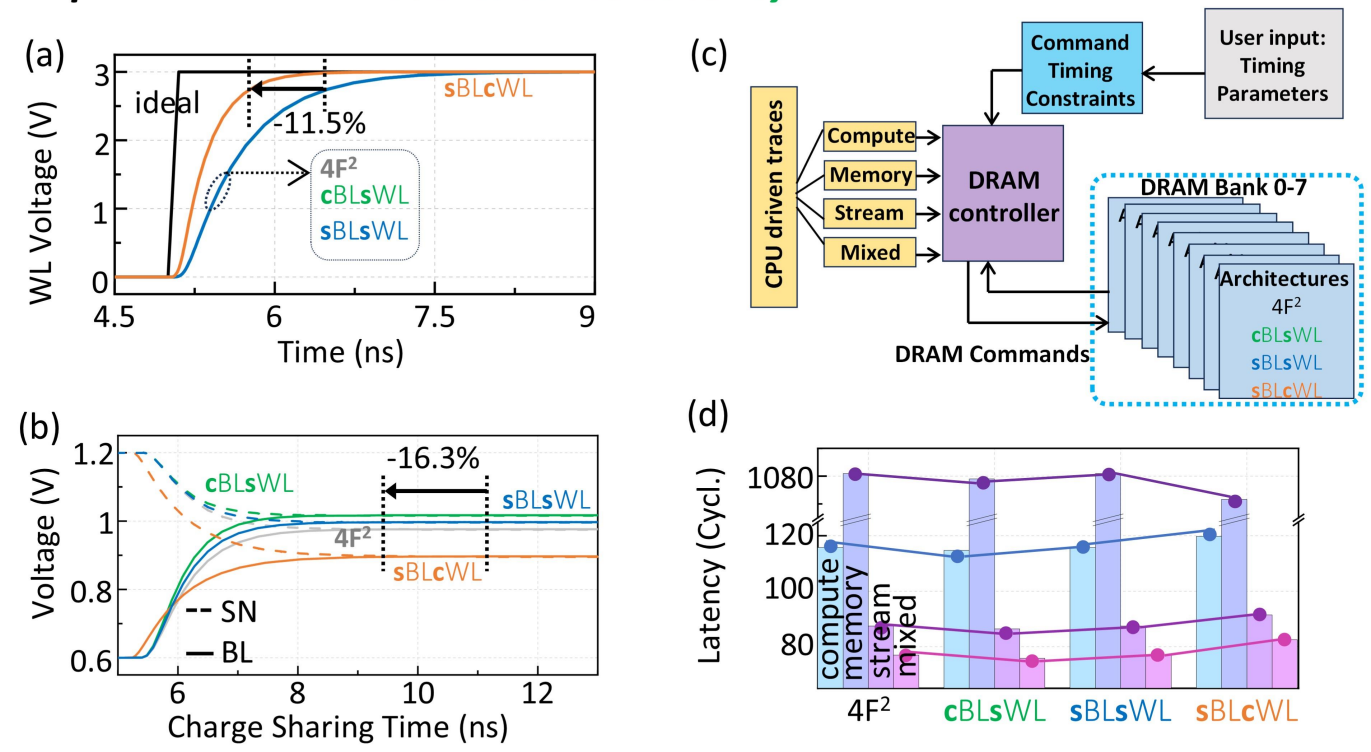

Fig. 9. (a) WL turn-on delay waveform. **s**BL**c**WL has the smallest delay for the smallest WL parasitics. (b) Charge sharing time comparison. (c) DRAM chip evaluation platform based on Ramulator. (d) Latency comparison under four different load traces. Overall, **c**BL**s**WL performs the best.

TABLE I. COMPARISON OF $4F^2$ WITH 3 FVCT ARCHITECTURES

| Architecture | Density | $C_{BL}$ | Noise | Read Time | Write Time | Latency |
|---|---|---|---|---|---|---|
| $4F^2$ | $4F^2$ | Normal | Bad | Middle | Normal | Middle |
| cBLsWL | $2F^2$ | Smallest | Good | Best | Normal | Best |
| sBLsWL | $2F^2$ | Normal | Best | Middle | Normal | Middle |
| sBLcWL | $2F^2$ | Normal | Bad | Worst | Best | Worst (Memory Driven Best) |

## IV. EXTENDIBILITY OF FVCT TO BG VCT AND $1.75F^2$

For other reported VCT DRAMs with extra back gate (BG) [3,17], our flip technology is also applicable by separating the BG on FS and BS with self-aligned BL as in Fig. 10(a).

Besides, by using hexagon close-packed cap placement [18], the $2F^2$ bitcell can be even scaled to $1.75F^2$ (Fig. 10(b)). From the evaluation in Figs. 10(c-f), although $1.75F^2$ has 9.5% worse WL delay than $2F^2$ due to the tighter WL pitch, it has 14% larger SM than $4F^2$ and performs well under the three traces for the shortest BL length except the memory trace.

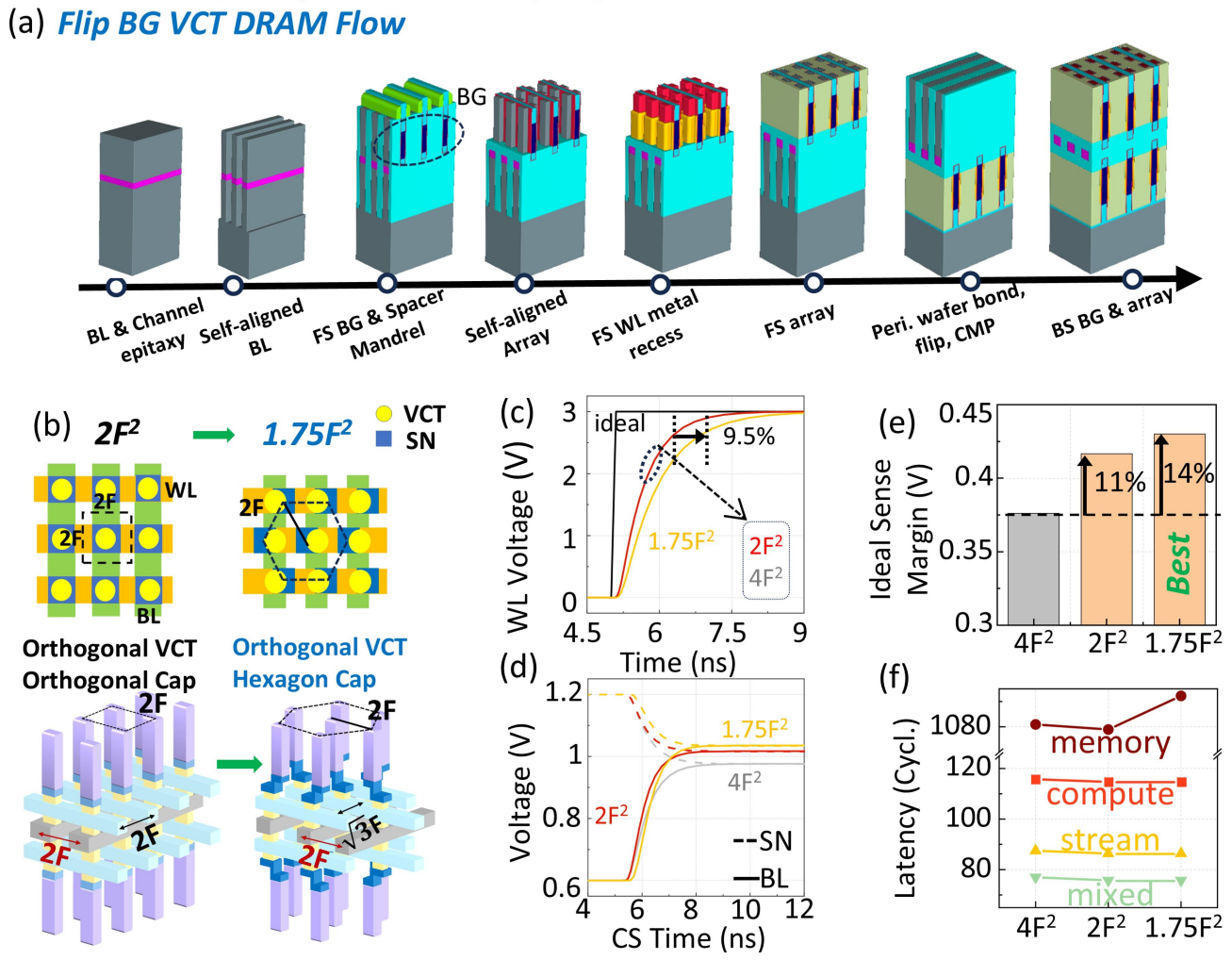

Fig. 10. (a) Flip technology extendibility to BG VCT, with self-aligned BL & split BG. (b) Hexagon cap + Orthogonal VCT for $1.75F^2$ design. (c-d) WL delay & CS time waveforms. (e) SM comparison, $1.75F^2$ is the best with the shortest BL length. (f) Latency from the chip evaluation.

## V. FLIP WL 3D-DRAM AND PERFORMANCE

As the multi-stack solution to DRAM, VBL 3D-DRAM also attracts wide attentions [19]. However, its HWL needs large-pitch staircases for connection, compromising the density gains from stacking. Even worse, the new stairless WL design on 3D-NAND [20] is not applicable in VBL 3D-DRAM [9]. To solve it, the novel FWL design places half of the WL staircases to the BS (Fig. 11(a)), cutting the staircase area by half as in Fig 11.(b). To accommodate this innovation, the HWL's peripheral circuits need to be adjusted accordingly for the array (Fig. 11(c)). In conventional 3D-DRAM, peripheral circuits are on one side of the wafer (Fig. 11(d)). However, for FWL 3D-DRAM, half of WL staircases are on the BS. To match this, sub-WL drivers (SWD) are placed on both FS and BS, saving peripheral area. Furthermore, cross bit line (CBL) strategy in 3D-NAND [21] can also be used to reduce BL SA area.

For a comprehensive analysis , a DTCO framework for FWL 3D-DRAM was also established, assuming VBL and HWL. For the WL & BL configurations, each WL connects 1024 cells horizontally while BL connects all the vertically stacked cells. From Fig. 12(a), the proposed FWL design greatly enhances the cell density by >20%, with a 15% reduced WL turn-on delay and 6.8% less CS time over the conventional 3D-DRAM as in Figs. 12(b-c). The benefits can be further validated by the chip level evaluation in Fig. 12(d) with greatly reduced chip latency.

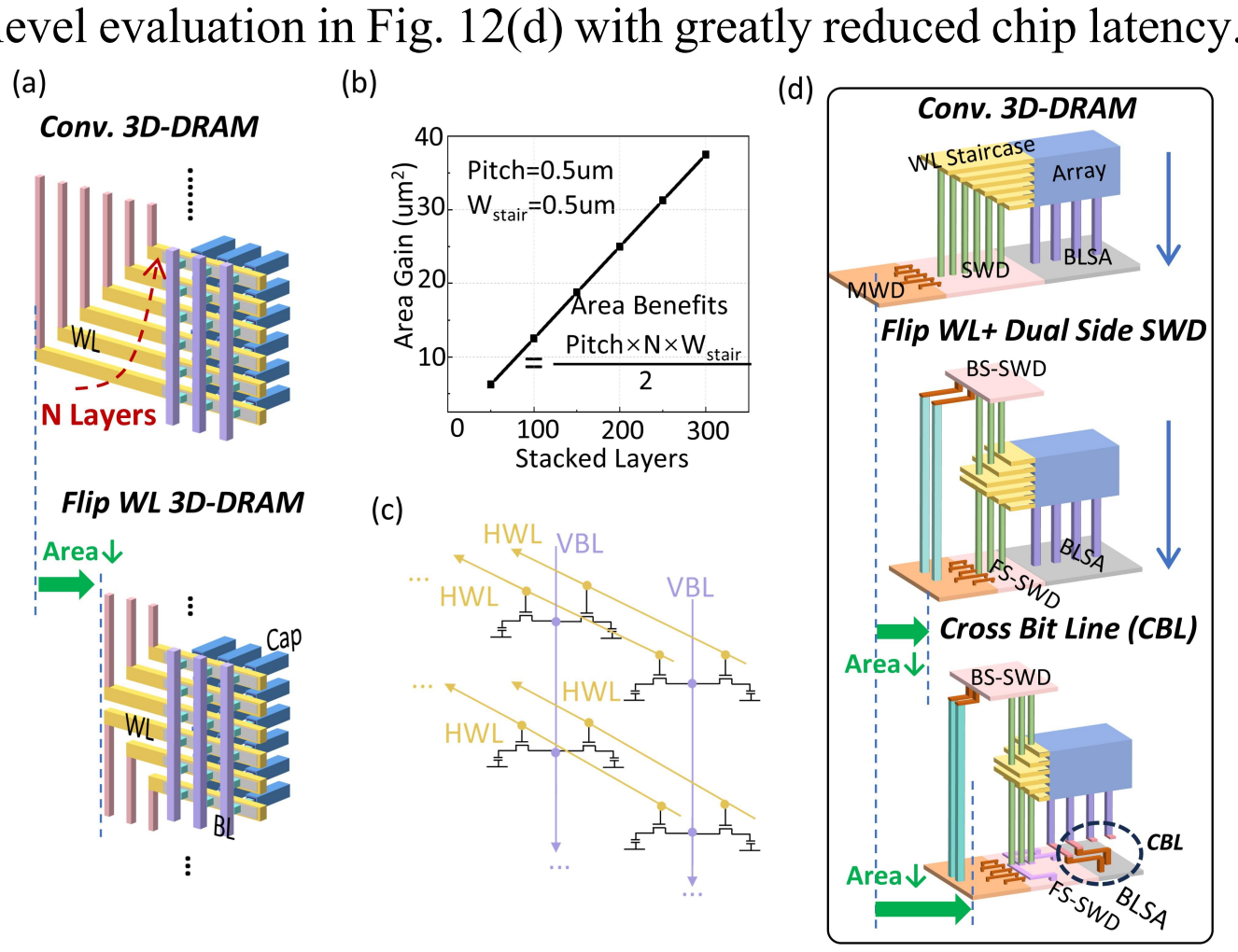

Fig. 11. (a) Flip WL 3D-DRAM puts half of WL staircases to the wafer backside, with great area benefits. (b) More layers stacked gives more area gains by FWL. (c) Circuits of VBL 3D-DRAM. (d) Peripheral circuit with dual-sided SWD & CBL for FWL 3D-DRAM.

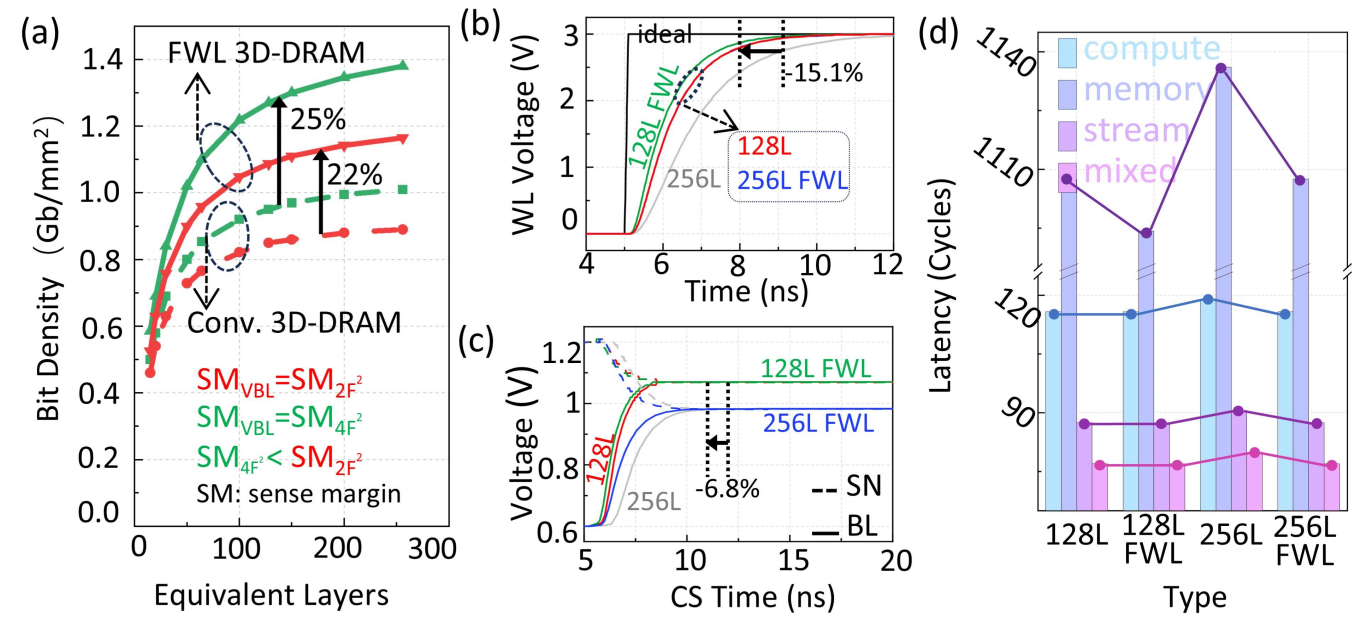

Fig. 12. (a) Bit density comparison under the same SM. (b-c) WL turn-on delay & CS time. (d) Latency benchmark at the chip level.

## VI. CONCLUSION

For here, a novel flip technology with process and architecture innovations is demonstrated on VCT DRAM & 3D-DRAM to further scale the DRAM beyond $4F^2$. Core processes on self-aligned $2F^2$ FVCT were developed and optimized for thermal and litho. challenges. 3 architectures of $2F^2$ FVCT were examined extensively from device to system DTCO. By flipping half of the WL staircases to BS, FWL 3D-DRAM with dual-sided SWDs improves both density and speed. This work provides new insights into DRAM roadmap.